\documentclass[a4paper,jou]{apa}
\usepackage[T1]{fontenc}
\usepackage[latin9]{inputenc}
\usepackage{color}
\definecolor{shadecolor}{rgb}{0.80078125, 0.80078125, 0.80078125}
\usepackage{framed}
\usepackage{amsmath}
\usepackage{amssymb}

\makeatletter

\pdfpageheight\paperheight
\pdfpagewidth\paperwidth

\helvetica
\author{Author} 
\affiliation{Affiliation} 

\note{Last updated: \today}

\journal{Ref.No: HG/GT.0313.01v1}
\copnum{Mar. 23th, 2013 -- Athens, Greece}

\acknowledgements{ Manuscript prepared by H.G. in LyX using the APA class.\\\indent Email: xgeorgio@di.uoa.gr --- http://xgeorgio.info}

\usepackage{multirow}

\usepackage{color}
\definecolor{CustomRed}{rgb}{.647,.129,.149}
\definecolor{CustomGreen}{rgb}{0.149,0.547,0.129}
\definecolor{CustomBlue}{rgb}{0.09,0.27,0.65}

\makeatother

\begin{document}

\title{\color{CustomRed}Feedback models and stability analysis of three
economic paradigms\normalcolor}

\rightheader{Ref.No: HG/GT.0313.01v1}

\author{Harris V. Georgiou}

\affiliation{\emph{Department of Informatics and Telecommunications,}\\
\emph{National \& Kapodistrian University of Athens, Greece.}}

\abstract{\emph{Abstract} --- In this paper, simple mathematical models from
Control Theory are applied to three very important economic paradigms,
namely (a) minimum wages in self-regulating markets, (b) market-versus-true
values and currency rates, and (c) government spending and taxation
levels. Analytical solutions are provided in all three paradigms and
some useful conclusions are drawn in terms of variable analysis. This
short study can be used as an example of how feedback models and stability
analysis can be applied as a guideline of 'proofs' in the context
of economic policies.}

\maketitle
\setcounter{page}{1}

\noindent {\LARGE T}HERE is a frequently-stated assertion that labor
cost is not the driving factor for production cost per unit, even
when the selling unit is not a product but some service. However,
in times of crisis and austerity, labor costs are almost always the
first (and usually the only) factor that is 'relaxed' to lower and
lower levels by enterprises, in an effort to keep the margin of profit
stable when selling rates decline. Some economists justify these policies
as the typical 'rule of thumb': when profits decline, the workers
will be paid less and less, until either the business recovers or
bankrupts. Others say that its is exactly the recipe of failure, since
underpaid workers will rarely work twice as hard to get the business
back on it feet - quite the opposite. 

Similarly, incentives for new private investments, e.g. low tax rates,
are often compared to public spending and the regulatory policies
are usually criticized as 'killers' for those incentives. However,
there is a definite link between changes in the investment flows and
the inherent gap between market value and true value of products and
services: excessively positive prospects cause a positive feedback
in new investments flow, while the exact opposite happens in times
of crises and markets downfall. This oscillatory feedback is (should
be) negated by an opposite feedback, usually realized under regulatory
policies (e.g. increasing tax rates and government spending), in order
to avoid the systemic risk of value 'bubbles' in both the market level
and a country's overall GDP change rate.

However, this is not the case in the real world; in fact, the exact
opposite happens. This is a 'paradox' of budgeting policies and government
spending that are based on false paradigms, hence the end result is
typically a self-reinforcing spiral: when things go well, hope and
money flow go up too; when things turn bad, more austerity policies
lead to the typical 'spiral of death', for a business or a whole country's
economy. This is a fundamental issue of high controversy among leading
economists and one that will be investigated in-depth under several
mathematical formulations in the next sections.

This is by no means a complete paper on economic policies nor in-depth
analysis of some of the most important issues in modern economics;
it is rather a simple, purely mathematical approach to three very
important paradigms, a short study that can be used as an example
of how feedback models and stability analysis from classic Control
Theory can be applied as a guideline of 'proofs' in the context of
economic policies.

The paper is organized as follows: The first section is a short formulation
for the problem of minimum wages and their importance in self-regulating
markets, under the scope of gain-cost analysis of private firms. Next,
the core issues of currency rates and differences between market value
and true (hidden) value are investigated under the scope of a first-order
feedback model, as well as a stability analysis with regard to private-versus-public
spending rates. Finally, government spending, public workers' wages
and the general taxation level are described in the context of a differential
equations model as mutually dependent variables of the same (economic)
system. Analytical solutions are provided in all three paradigms and
some useful conclusions are drawn in terms of variable analysis.

\section{\color{CustomRed}Minimum wages and self-regulating labor markets\normalcolor}

In order to investigate this 'labor cost shrinkage' dilemma and provide
some model-based guidelines, the driving force of every profit-focused
enterprise can be formulated in correspondence to the 'labor cost'
variable. Let $w=\sum p_{k}$ be the sum of the $p_{k}$ wages for
all $k$ workers, i.e., the enterprise's total labor cost per production
unit. Let $NC\left(w\right)$ be the total cost per production unit,
including labor cost, as a weighted sum of $n$ individual factors.
Without loss of generality, assume that the labor cost is at index
$i=1$, hence its weight in the total production cost is $\alpha_{1}$
and is singled out from the sum:

\begin{equation}
NC\left(w\right)=\underset{i=1}{\overset{n}{\sum}}\alpha_{i}\cdot z_{i}=\alpha_{1}\cdot w+\underset{i=2}{\overset{n}{\sum}}\alpha_{i}\cdot z_{i}\label{eq:NCw-def}
\end{equation}

The net profit $NP\left(w\right)$ per sold unit is proportional to
the difference between the maximum attainable market (selling) price
$MPr_{max}$ and the total cost per unit $NC\left(w\right)$, i.e.,
the maximum margin of profit, while inversely proportional to the
total labor cost $w$:

\begin{equation}
NP\left(w\right)\propto\frac{MPr_{max}-NC\left(w\right)}{w}\label{eq:NPw-def}
\end{equation}

Substituting (\ref{eq:NCw-def}) in (\ref{eq:NPw-def}) we have:

\begin{equation}
NP\left(w\right)=\frac{MP_{max}-\left(\alpha_{1}\cdot w+\underset{i=2}{\overset{n}{\sum}}\alpha_{i}\cdot z_{i}\right)}{w}=\frac{C}{w}-\alpha_{1}\label{eq:NPw-std}
\end{equation}

where $C=MPr_{max}-\underset{i=2}{\overset{n}{\sum}}\alpha_{i}\cdot z_{i}>0$
is a constant with respect to $w$ and $\alpha_{i}$ are typical convex
weighting factors, i.e.:$\underset{i=1}{\overset{n}{\sum}}\alpha_{i}=1\;,\;0\leq\alpha_{i}\leq1$.

In order to find the maximum attainable value for $NP\left(w\right)$,
the first- and second-order derivatives against $w$ must be calculated: 

\[
\frac{\partial NP(w)}{\partial w}=\frac{\partial}{\partial w}\left(\frac{C}{w}-\alpha_{1}\right)=\frac{-C}{w^{2}}\underset{C>0}{\overset{w\geq w_{0}\geq0}{\Longrightarrow}}\frac{\partial NP(w)}{\partial w}\leq0
\]

\[
\frac{\partial^{2}NP(w)}{\partial w^{2}}=\frac{\partial}{\partial w}\left(\frac{-C}{w^{2}}\right)=\frac{2C}{w^{3}}\underset{C>0}{\overset{w\geq w_{0}\geq0}{\Longrightarrow}}\frac{\partial^{2}NP(w)}{\partial w^{2}}\geq0
\]

In other words, the negative sign of the first-order derivative shows
(as expected) that $NP\left(w\right)$ decreases as $w$ increases
above some minimum allowable value $w_{0}$ (employing minimum wages
for the workers), while the positive sign of the second-order derivative
shows that the function curves upwards. These conditions are typical
to all functions $y=x^{p}$ where $k\leq1$, like in this case. Hence,
the function is maximized at the lower limit $w=w_{0}$: 

\begin{equation}
\begin{array}{c}
\left.\begin{array}{c}
\underset{w}{\max}NP\left(w\right)\\
s.t.:\: w\geq w_{0}\geq0
\end{array}\right\} \Rightarrow\underset{w\rightarrow w_{0}}{\lim}NP\left(w\right)=\\
=\underset{w\rightarrow w_{0}}{\lim}\left(\frac{C}{w}-\alpha_{1}\right)=\frac{C}{w_{0}}-\alpha_{1}=NP_{w_{0}}
\end{array}\label{eq:Wmin-0}
\end{equation}

where $NP_{w_{0}}$ is the maximum attainable net profit per sold
unit with regard to labor cost. Obviously, when there is no minimum-wage
limit, i.e., $w_{0}=0$, the net profit w.r.t. labor cost is sky-rocketed
to infinity: 

\begin{equation}
\begin{array}{c}
\left.\begin{array}{c}
\underset{w}{\max}NP\left(w\right)\\
s.t.:\: w\geq w_{0}=0
\end{array}\right\} \Rightarrow\underset{w\rightarrow0}{\lim}NP\left(w\right)=\\
=\underset{w\rightarrow0}{\lim}\left(\frac{C}{w}-\alpha_{1}\right)=+\infty=NP_{0}
\end{array}\label{eq:Wmin-inf}
\end{equation}

What (\ref{eq:Wmin-inf}) proves is nothing new:

\begin{center}
\parbox[c]{0.95\columnwidth}{%
\begin{shaded}%
The best labor wages policy, i.e., the one that provides the maximum
gain-per-salary ratio, is \emph{slavery} (free labor - no salaries
at all).\end{shaded}%
}
\par\end{center}

However, the most important conclusion from this result is the fact
that the minimum wage limit, as it is usually legislated by laws and
government policies, is actually not something that can be 'discovered'
by a totally free (unregulated) market. If salaries can go down to
zero, no enterprise has a serious incentive, profit-wise, to offer
a decent wage to anyone. In fact, even when minimum wage limits do
exist in a labor market, the difference between this lower threshold
and the actual mean value of offered wages is only marginal; it only
relies on the various enterprises' competitiveness over very few capable
candidates for many open positions. Of course, this is hardly the
case in the real world where the exact opposite is the rule, i.e.,
many highly qualified candidates have to compete for a limited number
of job openings.

\section{\color{CustomRed}A first-order feedback control model for stable
currency rates\normalcolor}

The starting point for the following model-based approach is the assertion
that there is a distinct difference between the true economy and the
currency used in it. That is, each commodity or labor effort has a
specific 'hidden' value that is \emph{invariant} with respect to a
currency that is used to 'translate' it into monetary value. This
assertion is valid for \emph{any} such currency, even those that are
bounded to a specific commodity, e.g. gold or silver, since no one
commodity can be used a the universal baseline for this evaluation:
a huge amount of gold is next to worthless when it can not buy food
or water if there are nowhere to be found. On the other hand, a single
worker can produce a specific amount of work (on average), just as
one apple tree can produce (on the same soil and climate) more or
less the same amount of apples, which in turn contain the same amount
of nutrients and calories. Hence, the \emph{market price} of a single
apple or a single man-hour of work, which is mainly a factor offer
and demand, is not a valid invariant metric of its \emph{true value}.

Furthermore, on monetary systems that are based on a representative
currency, i.e., some form of printed money or bonds, there is no inherent
link between the total volume of currency available to spend and the
total volume of commodities and labor to purchase - when there is
more money than goods, inflation occurs. More money means higher prices
for the exact same commodities, just like less commodities means higher
prices at the same total amount of money. It all depends on who controls
the total amount of currency available for spending and how this is
carefully balanced against the total amount of commodities to purchase.

\subsection{Basic model}

Let $V_{m}$ be the market price of the commodity, labor effort or
even the currency itself and let $V_{t}$ be the corresponding 'true'
value. These two values are correlated linearly by a parameter $\rho$
that corresponds to inflation and deflation effects in the economy:

\begin{equation}
V_{m}=\left(\rho+1\right)\cdot V_{t}\label{eq:Vm-r-Vt}
\end{equation}

When $\rho>0$ then inflation occurs, i.e., the market price is higher
than it should for some specific commodity or work effort unit, while
$\rho<0$ means deflation, i.e., the market price of the same commodity
gets devalued. Grouping inflation effects as $\varepsilon^{+}$ and
deflation effects as $\varepsilon^{-}$, the adjustment parameter
now becomes $\rho=\varepsilon^{+}-\varepsilon^{-}$.

Let us now consider these two factors, $\varepsilon^{+}$ and $\varepsilon^{-}$.
During inflation, there is a positive sign in the change rate of the
market price $V_{m}$ against the true value $V_{t}$, i.e.:

\begin{equation}
\varepsilon^{+}=c^{+}\cdot\frac{dV_{m}}{dV_{t}}\label{eq:e-dVmdVt-p}
\end{equation}

where $c^{+}\geq0$ is a constant. Likewise, during deflation, there
is a negative sign in the change rate of the market price $V_{m}$
against the true value $V_{t}$, i.e.:

\begin{equation}
\varepsilon^{-}=c^{-}\cdot\frac{dV_{m}}{dV_{t}}\label{eq:e-dVmdVt-n}
\end{equation}

where $c^{-}\geq0$ is a constant. Equations (\ref{eq:e-dVmdVt-p})
and (\ref{eq:e-dVmdVt-n}) essentially link the parameter $\rho$
in (\ref{eq:Vm-r-Vt}) with the change rate of the market price $V_{m}$
against the true value $V_{t}$. In other words, the inflation and
deflation effects are expressed as a function of the first derivative
of $V_{m}$ against $V_{t}$. 

The terms 'inflation' and 'deflation' here are somewhat misleading,
since they are usually linked to increasing and decreasing prices,
respectively, in the market. Here, 'inflation' is linked to its primary
definition, i.e., the increasing availability and flow of money in
the market, a situation that favors 'cheaper' currency, easier loans
and incentives for riskier investments. This means that when the money
flow increases, \emph{into} the economy (private investments, low
taxation levels, economy growth, etc), the incentives for moving even
more money into it increases, since profitable businesses are plenty.
On the other hand, 'deflation' here is also linked to its primary
definition, i.e., the decreasing availability and flow of money in
the market, a situation that results in 'expensive' currency, harder
loans and incentives for more conservative (or no) investments. This
means that when the money flow decreases, \emph{out of} the economy
(outgoing foreign exchange, high taxation levels, economy shrinking,
etc), the incentives for putting more and more money away from it
increases, since profitable investments are scarce. In other words,
equations (\ref{eq:e-dVmdVt-p}) and (\ref{eq:e-dVmdVt-n}) translate
the momentum of $V_{m}$ against $V_{t}$ into quantifiable feedback,
reinforcing (positive) or dampening (negative), according to the relation
between $c^{+}$ and $c^{-}$.

There is also a definite link between inflation/deflation and interest
rates in bank loans: during inflation, the governments and central
banks try to 'slow down' excessive loaning and credit card use by
raising the baseline for interest rates, while during deflation they
try to 'boost' the economy by lowering this baseline and thus enabling
more money flow in the market. Here, these 'correcting' actions are
essentially included in both $c^{+}$ and $c^{-}$, according to the
direction these authorities want to employ into (\ref{eq:e-dVmdVt-p})
and (\ref{eq:e-dVmdVt-n}) and, in the end, into (\ref{eq:Vm-r-Vt})
as well.

\subsection{Analytical solution}

The model described by (\ref{eq:Vm-r-Vt}) now becomes a differential
equation that is to be solved, i.e., fully defines $V_{m}$ as a function
of $V_{t}$. The following steps show how:

\[
\begin{array}{cc}
V_{m} & =\left(\rho+1\right)\cdot V_{t}=(\varepsilon^{+}-\varepsilon^{-}+1)\cdot V_{t}\\
 & =\left(\left(c^{+}-c^{-}\right)\cdot\frac{dV_{m}}{dV_{t}}+1\right)\cdot V_{t}\\
\Leftrightarrow & \left(c^{+}-c^{-}\right)\cdot V_{t}\cdot\frac{dV_{m}}{dV_{t}}-V_{m}+V_{t}=0\\
\Leftrightarrow & \left(c^{+}-c^{-}\right)\cdot x\cdot\frac{dy}{dx}-y+x=0\\
\Leftrightarrow & \frac{dy}{dx}+\left(\frac{-1}{c^{+}-c^{-}}\right)\cdot\frac{y}{x}+\left(\frac{1}{c^{+}-c^{-}}\right)=0
\end{array}
\]

\begin{equation}
\begin{array}{cc}
\Leftrightarrow & \frac{dy}{dx}=\beta\cdot\frac{y}{x}-\beta=F\left(\frac{y}{x}\right)\end{array}\label{eq:dydx-Fyx}
\end{equation}

where $y=V_{m}$, $x=V_{t}$ and $\beta=\frac{1}{c^{+}-c^{-}}$. The
last step in the previous sequence is essentially a typical transformation
into a well-known form of differential equations that can be solved
by substituting $z=\frac{y}{x}$ and calculating the integral:

\[
\begin{array}{cc}
 & \frac{dy}{dx}=\beta\cdot\frac{y}{x}-\beta=F\left(\frac{y}{x}\right)\\
\left(z=\frac{y}{x}\right) & \Rightarrow\ln x=\int\frac{dz}{F\left(z\right)-z}+\alpha\\
 & \ln x=\int\frac{dz}{F\left(z\right)-z}=\int\frac{d\left(\frac{y}{x}\right)}{\left(\beta\cdot\frac{y}{x}-\beta\right)-\left(\frac{y}{x}\right)}\\
 & =\int\frac{d\left(\frac{y}{x}\right)}{\frac{y}{x}\cdot\left(\beta-1\right)+\left(-\beta\right)}=\int\frac{dz}{z\cdot\left(\beta-1\right)+\left(-\beta\right)}\\
 & =\frac{1}{\left(\beta-1\right)}\cdot\ln\left(z\cdot\left(\beta-1\right)+\left(-\beta\right)\right)\\
\Leftrightarrow & \ln x=\frac{1}{\left(\beta-1\right)}\cdot\ln\left(z\cdot\left(\beta-1\right)+\left(-\beta\right)\right)\\
\Leftrightarrow & \ln x^{\left(\beta-1\right)}=\ln\left(\frac{y}{x}\cdot\left(\beta-1\right)+\left(-\beta\right)\right)\\
\Leftrightarrow & x^{\left(\beta-1\right)}=\frac{y}{x}\cdot\left(\beta-1\right)+\left(-\beta\right)\\
\Leftrightarrow & \left(\beta-1\right)\cdot y=x^{\beta}+\beta\cdot x
\end{array}
\]

\begin{equation}
\begin{array}{cc}
\Leftrightarrow & y=\left(\frac{1}{\beta-1}\right)\cdot x^{\beta}+\left(\frac{\beta}{\beta-1}\right)\cdot x\end{array}\label{eq:y-exp-x}
\end{equation}

Hence, from the final result an analytical formula of $V_{m}$ with
regard to $V_{t}$ becomes available:

\begin{equation}
V_{m}=\left(\frac{1}{\beta-1}\right)\cdot V_{t}^{\beta}+\left(\frac{\beta}{\beta-1}\right)\cdot V_{t}\qquad\beta=\frac{1}{c^{+}-c^{-}}\label{eq:Vm-exp-Vt}
\end{equation}

Note that (\ref{eq:Vm-exp-Vt}) is more or less the long-term expansion
of (\ref{eq:Vm-r-Vt}). That is, (\ref{eq:Vm-r-Vt}) is the 'instance'
definition of $V_{m}$ with regard to $V_{t}$ as a differential equation,
while in (\ref{eq:Vm-exp-Vt}) $V_{m}$ is defined only as a function
of $V_{t}$ (no differentials) and some constant parameters. This
analytical form is appropriate for calculating stability and feedback
factors as the result of these constant parameters and how they affect
the relation between $V_{m}$ and $V_{t}$.

As a verification step, one can calculate the differential term $\frac{dy}{dx}$
by its starting definition in (\ref{eq:dydx-Fyx}) and by its analytical
solution in (\ref{eq:y-exp-x}). From (\ref{eq:dydx-Fyx}) this calculation
gives:

\[
\begin{array}{cc}
\frac{dy}{dx} & =\beta\cdot\frac{y}{x}-\beta=\beta\cdot\left(\frac{x^{\beta}+\beta\cdot x}{\beta-1}\right)\cdot x^{-1}=...=\left(\frac{\beta}{\beta-1}\right)\cdot\left(x^{\beta-1}+\beta\right)\end{array}
\]

Similarly, from (\ref{eq:y-exp-x}) the same calculation gives:

\[
\begin{array}{cc}
\frac{dy}{dx} & =\left(\frac{1}{\beta-1}\right)\cdot\frac{d\left(x^{\beta}+\beta\cdot x\right)}{dx}=\frac{\beta\cdot x^{\beta-1}+\beta}{\beta-1}=\left(\frac{\beta}{\beta-1}\right)\cdot\left(x^{\beta-1}+\beta\right)\end{array}
\]

Hence, (\ref{eq:Vm-exp-Vt}) is a valid analytical solution of (\ref{eq:Vm-r-Vt}).

\subsection{Stability and feedback analysis}

Equation (\ref{eq:Vm-exp-Vt}) provides a full description for $V_{m}$
with regard to $V_{t}$ and the means to analyze its asymptotic behavior.
Since $\beta$ is the only parameter that includes all model configuration,
it is the main factor of interest here. Specifically, it is evident
that as $|\beta|$ increases, i.e., as the difference $|c^{+}-c^{-}|$
becomes smaller, $V_{m}$ exhibits larger exponent in $V_{t}$. This
means that $V_{m}$ either increases at higher rates (when $\beta>0$)
or drives the first term to zero (when $\beta<0$). Furthermore, as
$|\beta|$ becomes larger, the coefficient $\frac{\beta}{\beta-1}$
of the second term in (\ref{eq:Vm-exp-Vt}) approaches unity. In other
words, as $c^{+}$ approaches $c^{-}$, $V_{m}$ and $V_{t}$ become
directly proportional (not just linearly dependent). 

Combining these previous comments with respect to $\beta$, it is
clear that if $V_{m}$ is to be 'stabilized' against $V_{t}$, $\beta$
can be selected accordingly in order to diminish the first (higher-order)
term and reinforce the second (linear) term. This happens only when
$|\beta|\gg1$ and $\beta<0$, i.e., as $\beta\rightarrow-\infty$:

\begin{equation}
V_{m}=\left(\frac{1}{\beta-1}\right)\cdot V_{t}^{\beta}+\left(\frac{\beta}{\beta-1}\right)\cdot V_{t}\overset{\beta\rightarrow-\infty}{\longrightarrow}V_{t}^{+}\label{eq:Vm-lim-Vt}
\end{equation}

where $V_{t}^{+}$ means that it is approached from higher values
as $\beta$ becomes more and more negative. In other words:

\begin{equation}
\beta\rightarrow-\infty\Leftrightarrow c^{+}-c^{-}\rightarrow0^{-}\Longrightarrow V_{m}\rightarrow V_{t}^{+}\label{eq:Vm-lim-Vt-cc}
\end{equation}

What equation (\ref{eq:Vm-lim-Vt-cc}) says is pretty clear: 

\begin{center}
\parbox[c]{0.95\columnwidth}{%
\begin{shaded}%
If market prices of commodities and work effort are to be kept close
to their true values, the negative feedback factors (high taxes, high
interest rates, high government spending, etc) should \emph{closely
match} the positive feedback factors (low taxes, low interest rates,
strict government spending, etc).\end{shaded}%
}
\par\end{center}

This result is not something unexpected; keeping taxes and interest
rates high is the standard policy for slowing down a booming economy
into a 'controlled growth' state. This is necessary to avoid excessive
debt increase in both public and private sector, as well as decreasing
the incentives of 'bubbles' in stock markets. However, what is very
interesting is that the proper control policy is for the authorities
to \emph{counter match} the positive feedback with \emph{proportional}
negative feedback actions. In other words, government spending, taxation
levels and interest rates should \emph{always} increase/decrease in
proportion to private investments, creation of new businesses and
incoming flow of foreign capital. 

Unfortunately, the idea of a deliberate slowdown in the economy is
something that is often unthinkable for modern free trades and stock
markets - this is why (\ref{eq:Vm-lim-Vt}) and (\ref{eq:Vm-lim-Vt-cc})
also constitute a very realistic explanation of the various financial
crises, like the dot-com bubble of the late '90s or the 2008 house
market crash in USA: \emph{when negative feedback is not enforced,
the catastrophic deviation of $V_{m}$ from $V_{t}$ becomes a mathematical
certainty}.

\section{\color{CustomRed}Government spending and stable taxation level\normalcolor}

One of the most controversial issues in all economic models is the
acceptable amount of government spending for public services, infrastructure
and government salaries. In general, the amount of government budget
available for spending is directly proportional to the country's Gross
Domestic Product (GDP):

\begin{equation}
\left\{ GDP\right\} =C+I+G+\left(X-M\right)\thicksim S+P=W\label{eq:GDP-gen}
\end{equation}

\subsection{Basic model}

Equation (\ref{eq:GDP-gen}) is the typical calculation of GDP: $C$
is for consumption, $I$ is for investments and savings (domestic),
$G$ is for government spending and $\left(X-M\right)$ is the exports-imports
(net) balance. Reformulating these parameters, let $W$ be the GDP
portion attributed to salaries of workers in the public and the private
sector, i.e., $S$ and $P$ respectively. Public spending on infrastructure
can be reformulated as the weighted sum of $N$ factors, each contributing
$\alpha_{i}$ to the total spending:

\[
G_{s}=\sum\alpha_{i}g_{i}\quad i=1,...,N
\]

\[
\sum\alpha_{i}=1\quad0\leq\alpha_{i}\leq1
\]

Similarly, a 'public welfare' index can be calculated as a weighted
sum of $K$ factors, each contributing $\gamma_{i}$ to the index,
that affect availability of public services to the people:

\[
G_{w}=\sum\gamma_{k}\left(1-h_{k}\right)\quad k=1,...,K
\]

\[
\sum\gamma_{k}=1\quad0\leq\gamma_{k}\leq1
\]

where $h_{k}$ is the fraction of people without access to public
service $k$. Clearly, there is a link between $G_{s}$ and $G_{w}$,
i.e., $G_{w}\thicksim G_{s}$.

Let us now focus on the government budget that need to cover for public
workers' salaries $S$ and public spending $G_{s}$. Let $Q^{+}$be
the positive flow, which is essentially the sum of taxes on wages
to all workers (public and private sector), and let $Q^{-}$ be the
negative flow, which goes to $S$ and public spending $G_{s}$. If
$0\leq c\leq1$ is the balancing factor between salaries and infrastructure
in government spending and $p$ is the balancing factor between private
sector and public sector fractions in the total work force, then:

\[
Q^{+}=\left(S+P\right)\cdot t=\left(W\cdot\left(1-p\right)+W\cdot p\right)\cdot t=W\cdot t
\]

\[
Q^{-}=\left(1-c\right)\cdot\hat{S}+G_{s}\cdot c=\left(1-c\right)\cdot\left(1-t\right)\cdot S+G_{s}\cdot c
\]

For a long-term viable budget management without deficiencies, huge
reserving and external loans, then the positive/negative flows should
be roughly equal:

\[
Q^{+}\simeq Q^{-}\Leftrightarrow S\cdot t+P\cdot t\simeq\left(1-c\right)\cdot\left(1-t\right)\cdot S+G_{s}\cdot c
\]

\begin{equation}
Q^{+}-Q^{-}=W\cdot t-\left(1-c\right)\cdot\left(1-t\right)\cdot\left(1-p\right)\cdot W-G_{s}\cdot c=\varphi\label{eq:QQ-phi-def}
\end{equation}

where $\varphi$ is the instantaneous (annual) balance in the government
budget. The model presented above assumes perfect mechanisms for spending,
paying salaries and collecting taxes. For a more realistic calculation,
deficiency factors have to be introduced in all the major components
in (\ref{eq:QQ-phi-def}), i.e.:

\[
\hat{t}=t\cdot\left(1-\varepsilon_{t}\right)\;,\;0\leq\varepsilon_{t}\leq1\;,\;0\leq t\leq1
\]

\[
\hat{W}=W\cdot\left(1-\varepsilon_{w}\right)\;,\;0\leq\varepsilon_{w}\leq1
\]

\[
\hat{G}=G\cdot\left(1-\varepsilon_{g}\right)\;,\;0\leq\varepsilon_{g}\leq1
\]

\[
\hat{\varphi}=\varphi\cdot\left(1-\varepsilon_{\varphi}\right)\;,\;0\leq\varepsilon_{\varphi}\leq1
\]

Here, $\varepsilon_{t}$ stands for deficiency in collecting taxes,
$\varepsilon_{w}$ stands for deficiency in work effort (outsourced
workers), $\varepsilon_{g}$ stands for deficiency in constructing
and maintaining public services (corruption) and $\varepsilon_{\varphi}$
stands for deficiency due to inflation (domestic currency devaluation).
These adjusted components can be introduced directly into (\ref{eq:QQ-phi-def})
for proper calculations; however, for the shake of simplicity, the
simplified model of (\ref{eq:QQ-phi-def}) will be used as-is, since
this choice does not affect the analysis that follows.

\subsection{Analytical solution}

Returning now to (\ref{eq:GDP-gen}), the investments component $I$
can be expressed as a factor of $W$, meaning that the amount of money
available for domestic spending drives the incentives for more investments,
new businesses and foreign capitals:

\begin{equation}
I=\xi\cdot W\cdot\left(1-t\right)\cdot\left(1+\vartheta\right)\quad\vartheta,\xi\geq0\label{eq:Invest-W}
\end{equation}

where $\xi$ is the amount of available money $W\cdot\left(1-t\right)$
(after taxation) that goes into investments and $\vartheta$ is the
multiplier that is attributed to foreign capital that comes into the
domestic economy as investments too. Hence, the true annual change
in $W$ can now be stated as a function of $\varphi$ and $I$ as:

\begin{equation}
\begin{array}{ccc}
\triangle W & = & \varphi+I\\
 & = & W\cdot t-\left(1-c\right)\cdot\left(1-t\right)\cdot\left(1-p\right)\cdot W-G_{s}\cdot c\\
 & + & \xi\cdot W\cdot\left(1-t\right)\cdot\left(1+\vartheta\right)
\end{array}\label{eq:DW-phi-Inv}
\end{equation}

Equation (\ref{eq:DW-phi-Inv}) is a differential model that links
$W$ with its change rate and all the other factors. Since it is stated
in a discrete form (annual changes), it can be solved as a first-order
iterative equation, defining $a_{n}=W_{n}$ and substituting for all
the other factors:

\begin{equation}
a_{n+1}=a_{n}\cdot\left(A+B\right)+C\Leftrightarrow a_{n+1}-a_{n}\cdot\left(A+B\right)=C\label{eq:an-diff}
\end{equation}

\begin{equation}
A=t-\left(1-c\right)\cdot\left(1-t\right)\cdot\left(1-p\right)\label{eq:A-diff}
\end{equation}

\begin{equation}
B=\xi\cdot\left(1-t\right)\cdot\left(1+\vartheta\right)\label{eq:B-diff}
\end{equation}

\begin{equation}
C=-G_{s}\cdot c\label{eq:C-diff}
\end{equation}

Equation (\ref{eq:an-diff}) is solved by calculating the solution
of the corresponding homogeneous system $\left(C=0\right)$ and then
trying a solution similar to the right-hand side of the general equation.
The solution of the homogeneous system is:

\[
d_{n+1}-d_{n}\cdot\left(A+B\right)=0
\]

\begin{equation}
\lambda-\left(A+B\right)=0\Rightarrow\lambda=A+B\Rightarrow d_{n}=\left(A+B\right)^{n}\cdot d_{0}\label{eq:dn-diff}
\end{equation}

Since the right-hand side of (\ref{eq:an-diff}) is a zero-order polynomial,
a constant can be introduced as a solution to the general equation:

\begin{equation}
\hat{a}=b_{0}\Rightarrow b_{0}-\left(A+B\right)\cdot b_{0}=C\Leftrightarrow b_{0}=\frac{C}{1-\left(A+B\right)}\label{eq:b0-diff}
\end{equation}

Then, the complete solution of (\ref{eq:an-diff}) is the sum of the
partial solutions of (\ref{eq:dn-diff}) and (\ref{eq:b0-diff}),
i.e.:

\begin{equation}
a_{n}=d_{n}+\hat{a}=\left(A+B\right)^{n}\cdot d_{0}+b_{0}\label{eq:an-d0-b0}
\end{equation}

where $d_{0}$ is a constant that can be calculated directly by using
any value for $n$, i.e.:

\begin{equation}
n=0\rightarrow a_{0}=1\cdot d_{0}+b_{0}\Leftrightarrow d_{0}=a_{0}-b_{0}\label{eq:d0-value}
\end{equation}

Substituting (\ref{eq:b0-diff}) and (\ref{eq:d0-value}) into (\ref{eq:an-d0-b0}),
the final solution for $a_{n}$ becomes:

\[
\begin{array}{ccc}
a_{n} & = & \left(A+B\right)^{n}\cdot\left(a_{0}-b_{0}\right)+b_{0}\\
 & = & \left(A+B\right)^{n}\cdot\left(a_{0}-\frac{C}{1-\left(A+B\right)}\right)+\frac{C}{1-\left(A+B\right)}
\end{array}
\]

or in terms of the original $W$ parameter:

\begin{equation}
W_{n+1}=\left(A+B\right)^{n}\cdot\left(W_{0}-\frac{C}{1-\left(A+B\right)}\right)+\frac{C}{1-\left(A+B\right)}\label{eq:Wn-full}
\end{equation}

where $W_{0}$ is a constant corresponding to some starting value
for $W$. Hence, the total amount of money available as workers' salaries
(public and private sectors) is now expressed as a function of all
the other parameters of (\ref{eq:QQ-phi-def}) and (\ref{eq:Invest-W}).

\subsection{Stability and feedback analysis}

In order to evaluate the stability constraints for the model described
in (\ref{eq:Wn-full}), the most important factor is the base of the
exponent, i.e., $\left(A+B\right)$. The same result can be drawn
by applying the z-transformation to the original model in (\ref{eq:an-diff}):

\[
W_{n+1}-\left(A+B\right)\cdot W_{n}=C\overset{F(z)}{\longrightarrow}H\left(z\right)
\]

\begin{equation}
H\left(z\right)=\frac{C}{1-\left(A+B\right)\cdot z^{-1}}\longleftrightarrow h\left(n\right)=C\cdot\left(A+B\right)^{n}\cdot u\left(n\right)\label{eq:Hz-AB}
\end{equation}

It is clear from (\ref{eq:Hz-AB}) that, in order for the system to
be stable, the constraint $|A+B|\leq1$ needs to be true in all cases.
Let us now examine the case $A+B\leq1$ with regard to the taxation
level $t$, applying (\ref{eq:A-diff}) for $A$ and (\ref{eq:B-diff})
for $B$:

\[
\begin{array}{ccc}
 & A+B & \leq1\\
\Leftrightarrow & t-\left(1-c\right)\cdot\left(1-t\right)\cdot\left(1-p\right)+\xi\cdot\left(1-t\right)\cdot\left(1+\vartheta\right) & \leq1\\
\Leftrightarrow & t\cdot\left(1+\left(1-c\right)\cdot\left(1-p\right)-\xi\cdot\left(1+\vartheta\right)\right)\\
 & -\left(\left(1-c\right)\cdot\left(1-p\right)-\xi\cdot\left(1+\vartheta\right)\right) & \leq1\\
\Leftrightarrow & t\cdot\left(1+\tau\right)-\tau & \leq1
\end{array}
\]

where:

\begin{equation}
\tau=\left(1-c\right)\cdot\left(1-p\right)-\xi\cdot\left(1+\vartheta\right)\label{eq:tau-tt}
\end{equation}

and finally we get:

\begin{equation}
t\leq\frac{1+\tau}{1+\tau}=1\label{eq:tt-upper}
\end{equation}

Similarly, for the lower bound we get:

\[
\begin{array}{ccc}
 & A+B & \geq-1\\
\Leftrightarrow & t-\left(1-c\right)\cdot\left(1-t\right)\cdot\left(1-p\right)+\xi\cdot\left(1-t\right)\cdot\left(1+\vartheta\right) & \geq-1\\
\Leftrightarrow & t\cdot\left(1+\tau\right)-\tau & \geq-1
\end{array}
\]

which gives:

\begin{equation}
t\cdot\left(1+\tau\right)-\tau\geq-1\Leftrightarrow t\geq\frac{-1+\tau}{1+\tau}\label{eq:tt-lower}
\end{equation}

Combining (\ref{eq:tt-upper}) and (\ref{eq:tt-lower}), and since
$0\leq t\leq1$, we get the final range for 'stable' taxation level:

\begin{equation}
\max\left\{ 0,\frac{-1+\tau}{1+\tau}\right\} \leq t\leq1\label{eq:tt-st-range}
\end{equation}

Equation (\ref{eq:tt-st-range}) is essentially a range constraint
for $t$ and describes the stability conditions for $W$ with respect
to the taxation level. In practice, if the lower bound becomes positive,
this range is shrinking towards the upper bound, i.e., the taxation
levels are forced to be higher in order to maintain a stable budget
management. Using (\ref{eq:tau-tt}) this translates to:

\begin{equation}
\frac{-1+\tau}{1+\tau}>0\Leftrightarrow\tau>1\Leftrightarrow\left(1-c\right)\cdot\left(1-p\right)>1+\xi\cdot\left(1+\vartheta\right)\label{eq:tau-stability}
\end{equation}

The result stated by (\ref{eq:tau-stability}) is indeed a very interesting
one:

\begin{center}
\parbox[c]{0.95\columnwidth}{%
\begin{shaded}%
For a stable spending of government budget, the available range for
the taxation level is \emph{shrinking towards the higher limit} as
the total number of public workers and/or the spending weight (i.e.,
wages level) of their salaries becomes larger than the domestic and
foreign investments (incoming flow). \end{shaded}%
}
\par\end{center}

This statement per-se is completely expected, as government budget
comes from taxes and foreign capital investments (assuming no long-term
policies for debt deficiencies are allowed). However, if the total
government spending is assumed constant, (\ref{eq:tau-stability})
states that the system can also be stabilized by allocating more funds
to public infrastructure and services instead of public workers' wages.
In other words, \emph{budgets cuts (austerity) is not necessarily
the only solution available.} 

Equation (\ref{eq:tau-stability}) incorporates $c$ and $p$ as negative
terms, while $\xi$ and $\vartheta$ as positive ones, hence it is
fairly easy to come with another interesting result: Since $p$ corresponds
to the fraction of work force employed in the private sector, (\ref{eq:tau-stability})
implies that having an excessively large number of private sector
workers, with regard to true investments, essentially destabilizes
the system. This assertion can be explained by the fact that excessive
private worker force means excessive sum of salaries available for
spending, thus increased attraction of domestic and foreign funds
for new investments. This incentive essentially destabilizes the control
of $\triangle W$ in (\ref{eq:DW-phi-Inv}) and may cause a catastrophic
oscillation (market bubbles). Therefore, \emph{the proper stabilization
action is for the government to 'slow down' any excessive increases},
a result similar to the one stated at the end of the previous paradigm
(see previous section). This action is usually executed by employing
higher taxes and limiting incoming flow of investment funds - policies
that are usually considered unthinkable for modern free trades and
stock markets.

\section{\color{CustomRed}Conclusion\normalcolor}

In this paper, simple mathematical models from Control Theory were
applied to three very important economic paradigms, namely (a) minimum
wages in self-regulating markets, (b) market-versus-true values and
currency rates, and (c) government spending and taxation levels. 

The main conclusions are:
\begin{APAitemize}
\item The best labor wages policy, i.e., the one that provides the maximum
gain-per-salary ratio, is \emph{slavery} (free labor - no salaries
at all).
\item Even when minimum wage limits do exist in a labor market, the difference
between this lower threshold and the actual mean value of offered
wages is only marginal.
\item If market prices of commodities and work effort are to be kept close
to their true values, the negative feedback factors (high taxes, high
interest rates, high government spending, etc) should \emph{closely
match} the positive feedback factors (low taxes, low interest rates,
strict government spending, etc).
\item For stable economies, the proper control policy is for the authorities
(government) to \emph{counter match} the positive feedback (private
investments) with \emph{proportional} negative feedback actions.
\item For a stable spending of government budget, the available range for
the taxation level is \emph{shrinking towards the higher limit} as
the total number of public workers and/or the spending weight (i.e.,
wages level) of their salaries becomes larger than the domestic and
foreign investments (incoming flow).
\item If the total government spending is assumed constant, the budget can
also be stabilized by allocating more funds to public infrastructure
and services instead of public workers' wages (i.e. wages cuts is
not the only 'correcting' solution).
\end{APAitemize}
This short study that can be used as an example of how feedback models
and stability analysis can be applied as a guideline of 'proofs' in
the context of economic policies.

\nocite{*}
\bibliography{math-econ-proofs_apa-custom_HG-ver1a}

\end{document}